\documentclass[12pt]{article}
\usepackage[cp1251]{inputenc}
\usepackage[english]{babel}
\usepackage{amssymb,amsmath}
\usepackage{graphicx}

\title{ Differential cross sections $^4$He({$\gamma$,p})$^3$H and
$^4$He({$\gamma$,n})$^3$He reactions in the range of the photon
energies up to the threshold of the meson production}

\author{Yu.M.Arkatov, P.I.Vatset, V.I.Voloshchuk, V.A.Zolenko,\\
A.F.Khodyachikh, I.M.Prokhorets,  V.L.Marchenko, E.A.Vinokurov,\\
Yu.P.Lyakhno, S.I.Nagorny, Yu.A.Kasatkin, I.K.Kirichenko,\\
A.A.Zayats, V.N.Gur'ev}

\begin {document}

\maketitle

\begin{center} {\it National Science Center "Kharkov Institute of
Physics and Technology" \\ 61108, Kharkiv, Ukraine}
\end{center}

\begin{abstract}

Differential cross sections two-body ($\gamma,p$) and ($\gamma,n$)
reactions of the $^4$He nucleus disintegration were measured using
the bremsstrahlung beam of photons at the KIPT linac LEA-300 at
the maximum energy ${\rm E}_{\gamma }^{max}$ = 150~MeV Arkatov
{\it et al.} [Sov.J.Nucl.Phys.{\rm 10}, 639 (1970),
[Sov.J.Nucl.Phys.{\rm 19}, 598 (1974)]. The reaction products were
detected in a diffusion chamber placed in the magnetic field.
Later, this experiment was processed for the second time  Nagorny
{\it et al.} [Sov.J.Nucl.Phys. {\rm 53}, 228 (1991); Yad.Fiz. {\rm
53}, 365 (1991)]. At this the total number of handled events of
$^4{\rm He}$ disintegration was made up ${\sim }3\cdot {10}^4$ per
reaction channel. The differential cross sections were measured
with a 1~MeV step up to a photon energy of 45~MeV, and with a
greater step at higher energies. The step in the measurements of
the polar angle of nucleon emission was ${10}^{\circ}$ in the
c.m.s. Authors published data on differential cross sections only
at ${\rm E}_{\gamma }$=22.5, 27.5, 33.5, 40.5, 45, and 49 MeV
photon energies.

S.I.Nagorny, Yu.A.Kasatkin, I.K.Kirichenko, A.A.Zayats and
V.N.Gur'ev took part in the theoretical analysis of these data.
Lyakhno {\it et al.} [Nucl. Phys.{\rm A 781}, 306 (2007)] used
these data for combined analysis with data on cross section
asymmetry with linear polarized photons. Unfortunately, most
authors of these data already died. The mentioned above data are
kept only by Yu. Lyakhno. Unfortunately, part of the data of the
range of the photons energies 49$\le{\rm E}_{\gamma }\le$57 MeV
$^4$He({$\gamma$,p})$^3$H  reactions was lost. After the
discussion with the authors Yu. Lyakhno was assigned to put all
the saved data into the arXiv.

I set my sincere gratitude to authors of these data and also to
I.V. Dogyust for the help in the designing of this publication.

The cross sections are set in the $\mu$b/sr.


\end{abstract}

\vskip130pt

\begin{center}
{\bf T a b l e  1:  $^4$He({$\gamma$,n)$^3$He reaction}}
\begin{tabular}[t]{|c|c|c|c|c|c|c|}
\hline & \multicolumn{6}{|c|}{$E_{\gamma}$, MeV} \\ \cline{2-7}
$\theta_n^{\ast}$,deg & 22-23 & 23-24 &  24-25 & 25-26 & 26-27 &
27-28
\\ \hline 0-10 & 3.5 $\pm$ 2.4 & 4.5 $\pm$ 2.6 & 7.5$\pm$3.1 & 4$\pm$1.6 &
2.1$\pm$1.5 & 10.1$\pm$3.4 \\ \hline 10-20 & 6.9$\pm$3.5 &
9.0$\pm$3.7 & 21.2$\pm$5.2 &21.5$\pm$3.8 & 24.3$\pm$5.1 & 12.4$\pm$3.7 \\
\hline 20-30 & 20.7$\pm$6.0 & 39.2$\pm$7.7 & 42.5$\pm$7.3 &
53.1$\pm$6.0 & 57.0$\pm$7.8 & 49.6$\pm$7.5 \\ \hline 30-40 &
29.4$\pm$7.1 & 60.3$\pm$9.5 & 66.2$\pm$9.1 & 78.7$\pm$7.3 & 87.7$\pm$9.6 & 78.8$\pm$9.4 \\
\hline 40-50 & 51.9$\pm$9.5 & 87.5$\pm$11.5 & 97.5$\pm$11.0 &
113.0$\pm$ 8.7 & 95.1$\pm$10.0 & 98.0$\pm$10.5 \\ \hline 50-60 &
32.8$\pm$7.5 & 122.2$\pm$13.6 & 155.0$\pm$13.9 & 133.2$\pm$9.5 &
151.0$\pm$12.6 & 144.2$\pm$12.7 \\ \hline 60-70 & 72.6$\pm$11.2 &
93.5$\pm$11.9 &
156.2$\pm$14.0 & 160.1$\pm$10.4 & 172.2$\pm$13.5 & 147.5$\pm$12.9 \\
\hline 70-80 & 64.0$\pm$10.5 & 107.1$\pm$12.7 & 171.2$\pm$14.6 &
166.1$\pm$10.6 & 174.3$\pm$13.6 & 171.2$\pm$13.9 \\ \hline 80-90 &
69.1$\pm$10.9 & 137.2$\pm$14.4 & 177.5$\pm$14.9 & 193.7$\pm$11.4 &
226.0$\pm$15.5 & 190.3$\pm$14.6 \\ \hline 90-100 & 67.4$\pm$10.8 &
146.3$\pm$14.9 & 165.0$\pm$14.4 & 202.4$\pm$11.7 & 194.4$\pm$14.3
& 199.4$\pm$15.0 \\ \hline 100-110 & 64.0$\pm$10.5 &
114.6$\pm$13.1 &
141.2$\pm$13.3 & 174.9$\pm$10.8 & 192.2$\pm$14.3 & 181.3$\pm$14.3 \\
\hline 110-120 & 81.2$\pm$11.9 & 126.7$\pm$13.8 & 133.7$\pm$12.9 &
138.5$\pm$9.7 & 195.4$\pm$14.4 & 166.7$\pm$13.7 \\ \hline 120-130
& 53.6$\pm$9.6 & 110.1$\pm$12.9 & 103.7$\pm$11.4 & 131.8$\pm$9.4 &
152.1$\pm$12.7 & 149.9$\pm$13.0 \\ \hline 130-140 & 32.8$\pm$7.5 &
101.1$\pm$12.3 & 91.2$\pm$10.7 & 87.4$\pm$7.7 & 119.4$\pm$11.2 &
116.0$\pm$11.4 \\ \hline 140-150 & 22.5$\pm$6.2 & 43.7$\pm$8.1 &
71.2$\pm$9.4 & 55.8$\pm$6.1 & 88.7$\pm$9.7 & 69.8$\pm$8.9 \\
\hline 150-160 & 12.1$\pm$4.6 & 34.7$\pm$7.2 & 42.5$\pm$7.3 &
28.9$\pm$4.4 & 40.1$\pm$6.5 & 39.4$\pm$6.7 \\ \hline 160-170 &
3.5$\pm$2.4 &
16.6$\pm$5.0 & 13.7$\pm$4.1 & 21.5$\pm$3.8 & 22.2$\pm$4.8 & 13.5$\pm$3.9 \\
\hline 170-180 & 3.5$\pm$2.4 & 3.0$\pm$2.1 & 7.5$\pm$3.1 &
7.4$\pm$2.2 & 4.2$\pm$2.1 & 9.0$\pm$3.2 \\ \hline
\end{tabular}
\end{center}

\vskip10pt

\begin{center}
\begin{tabular}[t]{|c|c|c|c|c|c|c|}
\hline & \multicolumn{6}{|c|}{$E_{\gamma}$, MeV} \\ \cline{2-7}
$\theta_n^{\ast}$,deg & 28-29 & 29-30 &  30-31 & 31-32 & 32-33 &
33-34 \\ \hline 0-10 &   7.7$\pm$2.9 & 4.7$\pm$2.3 & 9.0$\pm$2.5 &
7.4$\pm$2.4 & 3.3$\pm$1.6 & 7.5$\pm$2.5 \\ \hline 10-20 &
13.2$\pm$3.8 &   22.1$\pm$5.1  &  20.2$\pm$3.7  &  17.8$\pm$3.6  &
16.3$\pm$3.7  &  15.9$\pm$3.7 \\ \hline 20-30 & 53.0$\pm$7.6  &
51.0$\pm$7.6  &  35.4$\pm$ 5.0  &  37.2$\pm$5.3  &  32.7$\pm$5.2
&  32.7$\pm$5.2 \\ \hline 30-40 & 67.3$\pm$8.6  &  75.7$\pm$9.4  &
57.7$\pm$6.3  &  54.3$\pm$6.3  &  42.5$\pm$5.9  &  41.9$\pm$5.9 \\
\hline 40-50 & 118.1$\pm$11.4 & 101.3$\pm$10.9 & 88.3$\pm$7.8  &
86.2$\pm$8.0  &  72.7$\pm$7.7  &  75.4$\pm$7.9\\ \hline 50-60 &
124.7$\pm$11.7 & 132.7$\pm$12.4 & 116.1$\pm$9.0  & 109.2$\pm$9.0 &
117.7$\pm$9.8  & 99.7$\pm$9.1 \\ \hline 60-70 & 155.6$\pm$13.1 &
165.3$\pm$13.9 & 139.7$\pm$9.9  & 143.4$\pm$10.3 & 125.1$\pm$ 10.1
& 132.3$\pm$10.5 \\ \hline 70-80 & 183.2$\pm$14.2 & 140.9$\pm$12.8
& 150.8$\pm$10.2 & 139.7$\pm$10.2 & 144.7$\pm$10.9 & 118.0$\pm$9.9
\\ \hline 80-90 & 174.4$\pm$13.9 & 199.1$\pm$15.2 & 169.6$\pm$10.9
& 142.7$\pm$10.3 & 151.2$\pm$11.1 & 149.1$\pm$11.2 \\ \hline
90-100 & 190.9$\pm$14.5 & 188.6$\pm$14.8  &  160.6$\pm$10.6 &
141.2$\pm$10.2 & 148.8$\pm$11.0 & 164.2$\pm$11.7 \\ \hline 100-110
& 170.0$\pm$13.7 & 181.6$\pm$14.5 & 164.0$\pm$10.7 &
155.3$\pm$10.7 & 135.7$\pm$10.5 & 170.0$\pm$11.9 \\ \hline 110-120
& 176.6$\pm$14.0 & 144.3$\pm$13.0 & 143.2$\pm$10.0 & 170.9
$\pm$11.3 & 144.7$\pm$10.9 & 119.8$\pm$10.0 \\ \hline 120-130 &
133.6$\pm$12.1 & 156.0$\pm$13.5 & 137.6$\pm$9.8  & 128.6$\pm$9.8
& 141.4$\pm$10.8 & 120.6$\pm$10.1 \\ \hline 130-140 &
109.3$\pm$11.0 & 95.5$\pm$10.5  & 103.6$\pm$8.5  & 98.1$\pm$8.5  &
94.8$\pm$8.8   & 92.1$\pm$8.8 \\ \hline 140-150 & 83.9$\pm$9.6  &
76.8$\pm$9.5  &  79.2$\pm$7.4  &  74.3$\pm$7.4   & 91.5$\pm$8.7  &
70.4$\pm$7.7 \\ \hline 150-160 & 48.6$\pm$7.3  &  50.1$\pm$7.6  &
58.4$\pm$6.4  &  44.6$\pm$5.8  &  57.2$\pm$6.8  &  42.7$\pm$6.0 \\
\hline 160-170 & 13.2$\pm$3.8  &  26.8$\pm$5,6  & 31.3$\pm$4.7   &
17.1$\pm$3.6   & 28.6$\pm$4.8  &  16.8$\pm$3.7\\ \hline 170-180 &
8.8$\pm$3.1  &  4.7$\pm$2.3 & 4.2$\pm$1.7 & 4.5$\pm$1.8 &
9.0$\pm$2.7 & 8.4$\pm$2.6 \\ \hline
\end{tabular}
\end{center}

\begin{center}
\begin{tabular}[t]{|c|c|c|c|c|c|c|}
\hline & \multicolumn{6}{|c|}{$E_{\gamma}$, MeV} \\ \cline{2-7}
$\theta_n^{\ast}$,deg & 34-35 & 35-36 &  36-37 & 37-38 & 38-39 &
39-40\\ \hline 0-10 &   12.3$\pm$3.6  &  12.7$\pm$3.3  &
4.9$\pm$2.2 &   6.1$\pm$2.5 & 8.9$\pm$3.1  &  2.5$\pm$1.8 \\
\hline 10-20 &  7.2$\pm$2.7 & 19.5$\pm$4.1  &  9.9$\pm$3.1 & 13.3$\pm$3.7 &   16.7$\pm$4.3 &   10.0$\pm$3.5 \\
\hline 20-30 & 36.0$\pm$6.1  &  38.9$\pm$5.7  &  25.6$\pm$5.0 &   25.5$\pm$5.1 &  27.8$\pm$5.6 & 22.4$\pm$5.3 \\
\hline 30-40 & 49.3$\pm$7.1   & 49.9$\pm$6.5  &  49.3$\pm$7.0  &  53.1$\pm$7.4 &   51.2$\pm$7.5 &   41.1$\pm$7.2\\
\hline 40-50 & 55.5$\pm$7.5  &  72.0$\pm$7.8  &  69.0$\pm$8.2  & 67.4$\pm$8.3  &  57.9$\pm$8.0  &  59.8$\pm$8.6\\
\hline 50-60 & 85.3$\pm$9.4  &  106.7$\pm$9.5  & 66.0$\pm$8.1  & 77.6$\pm$8.9   & 73.4$\pm$9.0  &  84.7$\pm$10.3 \\
\hline 60-70 & 108.9$\pm$10.6 & 111.7$\pm$9.7  & 79.8$\pm$8.9  & 82.7$\pm$9.2  &  75.7$\pm$9.2   & 76.0$\pm$9.7 \\
\hline 70-80 & 112.0$\pm$10.7 & 118.5$\pm$10.0 & 112.3$\pm$10.5 & 104.2$\pm$10.3 & 99.0$\pm$10.5  & 102.1$\pm$11.3 \\
\hline 80-90 & 113.0$\pm$10.8 & 148.1$\pm$11.2 & 113.3$\pm$10.6 & 105.2$\pm$10.4 & 110.2$\pm$11.1 & 88.4$\pm$10.5\\
\hline 90-100 & 117.1$\pm$11.0 & 149.8$\pm$11.3 & 112.3$\pm$10.5 & 94.0$\pm$9.8   & 104.6$\pm$10.8 & 105.8$\pm$11.5\\
\hline 100-110 & 139.7$\pm$12.0 & 140.5$\pm$10.9 & 122.2$\pm$11.0 & 126.6$\pm$11.4 & 112.4$\pm$11.2 & 105.8$\pm$11.5 \\
\hline 110-120 & 123.3$\pm$11.3 & 116.8$\pm$9.9  & 130.1$\pm$11.3 & 137.9$\pm$11.9 & 101.3$\pm$10.6 & 76.0$\pm$9.7 \\
\hline 120-130 & 108.9$\pm$10.6 & 129.5$\pm$10.5 & 103.5$\pm$10.1 & 103.2$\pm$10.3 & 91.2$\pm$10.1  & 73.5$\pm$9.6 \\
\hline 130-140 & 107.9$\pm$10.5 & 96.5$\pm$9.0   & 82.8$\pm$9.0 &  65.4$\pm$8.2 &   71.2$\pm$8.9  &  73.5$\pm$9.6 \\
\hline 140-150 & 66.8$\pm$8.3   & 66.9$\pm$7.5  & 69.0$\pm$8.2 &  61.3$\pm$7.9  &  62.3$\pm$8.3 &  53.5$\pm$8.2 \\
\hline 150-160 & 33.9$\pm$5.9  & 56.7$\pm$6.9  &  49.3$\pm$7.0  &  34.7$\pm$6.0  &  32.3$\pm$6.0  &  22.4$\pm$5.3 \\
\hline 160-170 & 20.5$\pm$4.6   & 27.1$\pm$4. 8  & 24.6$\pm$4.9  &  15.3$\pm$4.0  &  14.5$\pm$4.0  &  19.9$\pm$5.0\\
\hline 170-180 & 9.2$\pm$3.1 & 7.6$\pm$2.5  &  5.9$\pm$2.4 & 7.1$\pm$2.7 & 4.5$\pm$2.2 & 5.0$\pm$2.5\\
\hline
\end{tabular}
\end{center}

\vskip10pt

\begin{center}
\begin{tabular}[t]{|c|c|c|c|c|c|c|}
\hline & \multicolumn{6}{|c|}{$E_{\gamma}$, MeV} \\ \cline{2-7}
$\theta_n^{\ast}$,deg &40-41 & 41-42 &  42-43 & 43-44 & 44-46 &
46-48\\ \hline 0-10 &   5.9$\pm$2.9  &  4.5$\pm$1.6   &
3.6$\pm$1.6   &  5.0$\pm$1.8 & 0.8$\pm$0.8  &   4.2$\pm$1.9 \\
\hline 10-20 &  5.9$\pm$2.9 & 11.3$\pm$2.5  &  10.8$\pm$2.8  &  10.6$\pm$2.6   & 11.9$\pm$3.1  &  14.1$\pm$3.4 \\
\hline 20-30 & 32.4$\pm$6.9  &  25.4$\pm$3.8  &  27.4$\pm$4.4  &  22.5$\pm$3.8  &  16.7$\pm$3.6  &  19.1$\pm$4.0 \\
\hline 30-40 & 36.9$\pm$7.4  &  41.7$\pm$4.9  &  36.0$\pm$5.1  &  46.9$\pm$5.4  &  26.2$\pm$4.6  &  28.3$\pm$4.9\\
\hline 40-50 & 59.0$\pm$9.3  &  54.7$\pm$5.6  &  54.1$\pm$6.2  &  51.3$\pm$5.7  &  39.7$\pm$5.6  &  50.8$\pm$6.5\\
\hline 50-60 & 75.2$\pm$10.5  & 61.5$\pm$5.9  &  56.2$\pm$6.4  &  64.4$\pm$6.3  &  59.5$\pm$6.9  &  46.6$\pm$6.2 \\
\hline 60-70 & 59.0$\pm$9.3  &  78.4$\pm$6.7  &  62.7$\pm$6.7  &  75.1$\pm$6.9  &  42.0$\pm$5.8  &  57.4$\pm$6.9 \\
\hline 70-80 & 72.2$\pm$10.3  & 70.5$\pm$6.3  &  78.6$\pm$7.5  &  78.2$\pm$7.0  &  63.4$\pm$7.1  &  66.6$\pm$7.4 \\
\hline 80-90 & 89.9$\pm$11.5  & 70.5$\pm$6.3  &  77.1$\pm$7.5   & 102.0$\pm$8.0  &  60.3$\pm$ 6.9  &  69.9$\pm$7.6\\
\hline 90-100 & 92.9$\pm$11.7 &  70.5$\pm$6.3   & 65.6$\pm$6.9  &  73.8$\pm$6.8  &  58.7$\pm$6.8  &  66.6$\pm$7.4\\
\hline 100-110 & 60.4$\pm$9.4  &  77.3$\pm$6.6  &  70.6$\pm$7.1  &  92.6$\pm$7.6  &  66.6$\pm$7.3  &  46.6$\pm$6.2 \\
\hline 110-120 & 92.9$\pm$11.7  & 76.1$\pm$6.6  &  67.0$\pm$7.0  &  75.1$\pm$6.9  &  53.1$\pm$6.5  &  57.4$\pm$6.9 \\
\hline 120-130 & 84.0$\pm$11.1  & 63.7$\pm$6.0  &  53.3$\pm$6.2  &  75.1$\pm$6.9  &  39.7$\pm$5.6  &  53.3$\pm$6.7\\
\hline 130-140 & 35.4$\pm$7.2  &  63.7$\pm$6.0  &  48.3$\pm$5.9  & 50.7$\pm$5.6  &  35.7$\pm$5.3  & 32.5$\pm$5.2 \\
\hline 140-150 & 31.0$\pm$6.8  &  47.4$\pm$5.2  & 33.9$\pm$4.9  &  45.0$\pm$5.3  &  27.0$\pm$4.6  &  26.6$\pm$4.7 \\
\hline 150-160 & 25.1$\pm$6.1  &  27.1$\pm$3.9  &  25.2$\pm$4.3  &  19.4$\pm$3.5  &  15.1$\pm$3 5  &  18.3$\pm$3.9 \\
\hline 160-170 & 11.8$\pm$4.2  &  10.2$\pm$2.4  & 8.6$\pm$2.5 & 9.4$\pm$2.4 & 4.8$\pm$1.9 & 12.5$\pm$3.2\\
\hline 170-180 & 2.9$\pm$2.1 & 3.4$\pm$1.4 & 5.8$\pm$2.0 & 6.3$\pm$2.0 & 0.8$\pm$0.8 & 5.0$\pm$2.0\\
\hline
\end{tabular}
\end{center}

\begin{center}
\begin{tabular}[t]{|c|c|c|c|c|c|c|}
\hline & \multicolumn{6}{|c|}{$E_{\gamma}$, MeV} \\ \cline{2-7}
$\theta_n^{\ast}$,deg &48-50 & 50-52 &  52-54 & 54-56 & 56-58 &
58-60\\ \hline 0-10 &   2.1$\pm$1.5  &   3.6$\pm$1.8 & 1.9$\pm$1.4
& 4.8$\pm$2.1 & 0.0$\pm$0.8  &  0.0$\pm$1.0 \\
\hline 10-20 &  6.2$\pm$2.5 & 2.7$\pm$1.6 & 3.9$\pm$1.9 & 8.6$\pm$2.9 & 3.2$\pm$1.6 & 7.0$\pm$2.7 \\
\hline 20-30 & 22.6$\pm$4.8  &  21.5$\pm$4.4  &  11.7$\pm$3.4  &  12.4$\pm$3.4   & 11.3$\pm$3.0 &   9.0$\pm$3.0 \\
\hline 30-40 & 28.8$\pm$5.4  &  34.9$\pm$5.6  &  13.6$\pm$3.6  &  10.5$\pm$3.2  &  22.6$\pm$4.3  &  16.1$\pm$4.0\\
\hline 40-50 & 48.3$\pm$7.1  &  36.7$\pm$5.7  &  35.9$\pm$5.9  &  27.6$\pm$5.1   & 24.2$\pm$4.4   & 20.1$\pm$4.5\\
\hline 50-60 & 42.2$\pm$6.6   & 51.0$\pm$6.8  &  24.3$\pm$4.9  &  41.9$\pm$6.3   & 34.7$\pm$5.3  &  29.1$\pm$5.4 \\
\hline 60-70 & 45.3$\pm$6.8  &  40.3$\pm$6.0  &  28.2$\pm$5.2  &  31.5$\pm$5.5  &  33.0$\pm$5.2  &  23.1$\pm$4.8 \\
\hline 70-80 & 50.4$\pm$7.2  &  42.1$\pm$6.1   & 40.8$\pm$6.3   & 35.3$\pm$5.8  &  33.9$\pm$5.2  &  35.2$\pm$5.9 \\
\hline 80-90 & 55.5$\pm$7.6  &  47.5$\pm$6.5   & 41.8$\pm$6.4  &  41.9$\pm$6.3  &  26.6$\pm$4.6   & 34.2$\pm$5.9\\
\hline 90-100 & 50.4$\pm$7.2  &  50.1$\pm$6.7  &  34.0$\pm$5.7  &  34.3$\pm$5.7  &  27.4$\pm$4.7  &  37.2$\pm$6.1\\
\hline 100-110 & 50.4$\pm$7.2  &  40.3$\pm$6.0  &  30.1$\pm$5.4  &  39.1$\pm$6.1  &  29.8$\pm$4.9  &  29.1$\pm$5.4 \\
\hline 110-120 & 59.6$\pm$7.8  &  41.2$\pm$6.1  &  35.0$\pm$5.8  &  35.3$\pm$5.8  &  25.8$\pm$4.6   & 46.2$\pm$6.8 \\
\hline 120-130 & 40.1$\pm$6.4  &  39.4$\pm$5.9  &  29.1$\pm$5.3  &  29.6$\pm$5.3  &  18.5$\pm$3.9  &  25.1$\pm$5.0\\
\hline 130-140 & 30.0$\pm$5.6  &  41.2$\pm$6.1  &  22.3$\pm$4.7  & 26.7$\pm$5.0   & 25.0$\pm$4.5  &  25.1$\pm$5.0 \\
\hline 140-150 & 16.5$\pm$4.1  &  20.6$\pm$4.3   & 17.5$\pm$4.1  & 23.8$\pm$4.8  &  15.3$\pm$3.5  &  18.1$\pm$4.3 \\
\hline 150-160 & 11.3$\pm$3.4  &  14.3$\pm$3.6  &  11.7$\pm$3.4  &  17.2$\pm$4.0  &  10.5$\pm$2.9  &  9.0$\pm$3.0 \\
\hline 160-170 & 4.1$\pm$2.1 & 6.3$\pm$2.4 & 5.8$\pm$2.4 & 2.9$\pm$1.7 & 4.8$\pm$2.0 & 7.0$\pm$2.7 \\
\hline 170-180 & 3.1$\pm$1.8 & 0.9$\pm$0.9 & 1.0$\pm$1.0 & 3.8$\pm$1.9 & 1.6$\pm$1.1 & 1.0$\pm$1.0\\
\hline
\end{tabular}
\end{center}

\vskip10pt

\begin{center}
\begin{tabular}[t]{|c|c|c|c|c|c|c|}
\hline & \multicolumn{6}{|c|}{$E_{\gamma}$, MeV} \\ \cline{2-7}
$\theta_n^{\ast}$,deg & 60-62 & 62-64 &  64-66& 66-68 & 68-70 &
70-72\\ \hline 0-10 &   1.1$\pm$0.8 & 0.0$\pm$0.7 &  0.8$\pm$0.8 &
0.7$\pm$0.7 & 1.2$\pm$0.9 & 0.6$\pm$0.6 \\
\hline 10-20 &  6.7$\pm$1.9 & 7.9$\pm$2.4 & 0.8$\pm$0.8 & 3.9$\pm$1.6 & 1.2$\pm$0.9 & 7.0$\pm$2.1 \\
\hline 20-30 & 12.2$\pm$2.6  &  7.9$\pm$2.4 & 8.3$\pm$2.6 & 11.2$\pm$2.7   & 10.5$\pm$2.5  &  7.0$\pm$2.1 \\
\hline 30-40 & 20.5$\pm$3.4  &  15.0$\pm$3.3  &  16.7$\pm$3.7  &  11.8$\pm$2.8  &  11.7$\pm$2.7  &  10.8$\pm$2.6 \\
\hline 40-50 & 27.8$\pm$3.9   & 21.4$\pm$3.9  &  15.8$\pm$3.6  &  10.5$\pm$2.6   & 12.9$\pm$2.8  &  12.7$\pm$2.8 \\
\hline 50-60 & 31.7$\pm$4.2  &  21.4$\pm$3.9  &  35.9$\pm$5.5  &  13.8$\pm$3.0  &  16.0$\pm$3.1  &  12.7$\pm$2.8 \\
\hline 60-70 & 26.1$\pm$3.8  &  19.3$\pm$3.7  &  26.7$\pm$4.7  &  17.1$\pm$3.4   &  20.3$\pm$3.5  &  23.4$\pm$3.9 \\
\hline 70-80 & 25.5$\pm$3.8  &  25.0$\pm$4.2   & 32.5$\pm$5.2  &  25.0$\pm$4.1  &  16.0$\pm$3.1  &  29.8$\pm$4.3 \\
\hline 80-90 & 27.2$\pm$3.9  &  23.6$\pm$4.1  &  29.2$\pm$4.9  &  29.6$\pm$4.4   & 24.6$\pm$3.9  &  21.5$\pm$3.7\\
\hline 90-100 & 28.3$\pm$4.0  &  32.1$\pm$4.8  &  32.5$\pm$5.2  &  25.7$\pm$4.1  &  26.5$\pm$4.0  &  18.4$\pm$3.4\\
\hline 100-110 & 26.7$\pm$3.8  &  27.1$\pm$4.4  &  32.5$\pm$5.2  &  20.4$\pm$3.7  &  17.2$\pm$3.3  &  22.2$\pm$3.7 \\
\hline 110-120 & 27.2$\pm$3.9   & 25.7$\pm$4.3  &  15.8$\pm$3.6   & 21.7$\pm$3.8  &  18.5$\pm$3.4  &  20.3$\pm$3.6\\
\hline 120-130 & 23.9$\pm$3.6  &  18.6$\pm$3.6  &  21.7$\pm$4.3 &  17.1$\pm$3.4  & 17.2$\pm$3.3  &  12.7$\pm$2.8 \\
\hline 130-140 & 21.7$\pm$3.5 &  15.7$\pm$3.3   & 14.2$\pm$3.4  &  13.8$\pm$3.0  &  16.6$\pm$3.2  &  13.9$\pm$3.0 \\
\hline 140-150 & 16.1$\pm$3.0  &  9.3$\pm$2.6 & 10.8$\pm$3.0  &  9.9$\pm$2.5 & 8.0$\pm$2.2 & 8.9$\pm$2.4 \\
\hline 150-160 & 6.1$\pm$1.8 & 9.3$\pm$2.6 & 7.5$\pm$2.5 & 5.3$\pm$1.9 & 4.3$\pm$1.6 & 7.0$\pm$2.1 \\
\hline 160-170 & 5.0$\pm$1.7 & 4.3$\pm$1.7 & 4.2$\pm$1.9 & 2.6$\pm$1.3 & 4.3$\pm$1.6 & 3.8$\pm$1.6 \\
\hline 170-180 & 0.6$\pm$0.6 & 2.1$\pm$1.2 & 2.5$\pm$1.4 & 3.3$\pm$1.5 & 0.6$\pm$0.6 & 1.9$\pm$1.1 \\
\hline
\end{tabular}
\end{center}

\begin{center}
\begin{tabular}[t]{|c|c|c|c|c|c|c|}
\hline & \multicolumn{6}{|c|}{$E_{\gamma}$, MeV} \\ \cline{2-7}
$\theta_n^{\ast}$,deg & 72-74 & 74-76 &  76-78 & 78-80 & 80-85 &
85-90\\ \hline 0-10 &   1.9$\pm$1.1 & 0.5$\pm$0.5 & 0.9$\pm$0.6 &
0.5$\pm$0.5 & 0.4$\pm$0.4 & 1.5$\pm$0.8 \\ \hline 10-20 &
2.5$\pm$1.2 & 4.6$\pm$1.5 & 3.4$\pm$1.2 & 5.3$\pm$1.7 &
3.7$\pm$1.2  & 1.5$\pm$0.8 \\ \hline 20-30 & 5.0$\pm$1.8 &
7.2$\pm$1.9 & 7.3$\pm$1.8 & 8.6$\pm$2.1 & 8.1$\pm$1.7 &
5.3$\pm$1.4 \\ \hline 30-40 & 15.0$\pm$3.1  &  10.3$\pm$2.3  &
6.9$\pm$1.7 & 11.2$\pm$2.4  &  7.0$\pm$1.6 & 5.3$\pm$1.4\\ \hline
40-50 & 20.6$\pm$3.6  &   9.2$\pm$2.2 & 13.4$\pm$2.4   &
15.5$\pm$2.9  &  11.7$\pm$2.1  &  10.6$\pm$2.0\\ \hline 50-60 &
15.6$\pm$3.1  &  20.5$\pm$3.2  &  15.1$\pm$2.5  &  14.4$\pm$2.8  &
14.7$\pm$2.3 &  14.8$\pm$2.4 \\ \hline 60-70 & 19.4$\pm$3.5  &
14.9$\pm$2.8  &  16.4$\pm$2.7  &  20.8$\pm$3.3  &  12.5$\pm$2.1  &
16.3$\pm$2.5 \\ \hline 70-80 & 20.0$\pm$3.5  &  15.4$\pm$2.8   &
19.4$\pm$2.9  &  21.9$\pm$3.4  &  13.6$\pm$2.2  &  14.0$\pm$2.3 \\
\hline 80-90 & 16.9$\pm$3.2  &  16.4$\pm$2.9  &  13.8$\pm$2.4  &
13.9$\pm$2.7  &  9.5$\pm$1.9 & 12.5$\pm$2.2 \\ \hline 90-100 &
25.6$\pm$4.0  &  13.9$\pm$2.7  &  14.2$\pm$2.5  &  16.0$\pm$2.9  &
10.6$\pm$2.0  &  10.6$\pm$2.0\\ \hline 100-110 & 20.0$\pm$3.5  &
13.9$\pm$2.7  &  11.2$\pm$2.2  &  17.6$\pm$3.1  &  8.4$\pm$1.8 &
10.2$\pm$2.0 \\ \hline 110-120 & 18.1$\pm$3.4  &  13.4$\pm$2.6  &
9.9$\pm$2.1 & 15.0$\pm$2.8  &  9.5$\pm$1.9 & 7.2$\pm$1.7 \\ \hline
120-130 & 14.4$\pm$3.0  &  15.4$\pm$2.8  &  12.5$\pm$2.3  &
11.2$\pm$2.4 & 8.4$\pm$1.8  &  10.6$\pm$2.0 \\ \hline 130-140 &
13.7$\pm$2.9   & 11.8$\pm$2.5  &  5.2$\pm$1.5 & 9.6$\pm$2.3 &
7.0$\pm$1.6  & 9.1$\pm$1.9 \\ \hline 140-150 & 15.0$\pm$3.1  &
7.7$\pm$2.0 & 5.2$\pm$1.5 & 8.6$\pm$2.1 & 7.0$\pm$1.6  &
4.2$\pm$1.3 \\ \hline 150-160 & 5.6$\pm$1.9 & 7.7$\pm$2.0 &
1.7$\pm$0.9 & 3.2$\pm$1.3 & 3.3$\pm$1.1  &  3.0$\pm$1.1 \\ \hline
160-170 & 1.2$\pm$0.9 & 5.7$\pm$1.7 & 2.6$\pm$1.1 & 3.7$\pm$1.4 &
0.0$\pm$0.4 & 2.3$\pm$0.9 \\ \hline 170-180 & 0.6$\pm$0.6 &
1.5$\pm$0.9 & 0.4$\pm$0.4 & 0.5$\pm$0.5 & 0.4$\pm$0.4 &
1.1$\pm$0.7 \\ \hline
\end{tabular}
\end{center}

\vskip10pt

\begin{center}
\begin{tabular}[t]{|c|c|c|c|c|c|c|}
\hline & \multicolumn{6}{|c|}{$E_{\gamma}$, MeV} \\ \cline{2-7}
$\theta_n^{\ast}$,deg & 90-95 & 95-100 &  100-110 & 110-120 &
120-130 & 130-140 \\
\hline 0-10 &   0.0$\pm$0.3 & 1.6$\pm$0.6 &0.2$\pm$0.2 &
0.2$\pm$0.1 & 0.2$\pm$0.1 & 0.2$\pm$0.1 \\
\hline 10-20 &  3.0$\pm$1.0 & 1.6$\pm$0.6 & 1.3$\pm$0.4 &
0.5$\pm$0.2 & 0.6$\pm$0.2 & 0.7$\pm$0.2 \\
\hline 20-30 & 6.7$\pm$1.5 & 4.5$\pm$1.0 & 2.3$\pm$0.5 &
1.5$\pm$0.4 & 1.8$\pm$0.4 & 1.0$\pm$0.2 \\
\hline 30-40 & 10.0$\pm$1.8 &   6.5$\pm$1.2 & 3.8$\pm$0.7 &
2.8$\pm$0.5 & 2.7$\pm$0.4 & 1.5$\pm$0.3\\
\hline 40-50 & 10.0$\pm$1.8 &   8.3$\pm$1.4 & 4.7$\pm$0.7 &
2.6$\pm$0.5 & 2.4$\pm$0.4 & 2.2$\pm$0.3\\
\hline 50-60 & 9.0$\pm$1.7 & 9.0$\pm$1.4 & 4.2$\pm$0.7  &
4.3$\pm$0.6 & 3.8$\pm$0.5 & 2.4$\pm$0.4 \\
\hline 60-70 & 13.0$\pm$2.1  &  9.4$\pm$1.5  &   4.8$\pm$0.8 &
3.5$\pm$0.6 & 3.0$\pm$0.5 & 1.6$\pm$0.3 \\
\hline 70-80 & 15.0$\pm$2.2  &   8.5$\pm$1.4 & 3.7$\pm$0.7 &
5.1$\pm$0.7 & 2.7$\pm$0.4 & 2.0$\pm$0.3 \\
\hline 80-90 & 16.0$\pm$2.3  &  7.4$\pm$1.3 & 4.3$\pm$0.7 &
2.9$\pm$0.5 & 2.1$\pm$0.4 & 1.7$\pm$0.3 \\
\hline 90-100 & 7.3$\pm$1.6 & 5.8$\pm$1.1 & 3.9$\pm$0.7  &
2.9$\pm$0.5 & 1.9$\pm$0.4 & 1.5$\pm$0.3\\
\hline 100-110 & 8.3$\pm$1.7 & 6.3$\pm$1.2  &   1.9$\pm$0.5 &
2.2$\pm$0.4 & 1.2$\pm$0.3 & 0.9$\pm$0.2 \\
\hline 110-120 & 7.7$\pm$1.6 & 4.3$\pm$1.0 & 3.0$\pm$0.6 &
1.1$\pm$0.3 & 1.3$\pm$0.3 & 0.9$\pm$0.2 \\
\hline 120-130 & 6.0$\pm$1.4 & 3.6$\pm$0.9 & 1.4$\pm$0.4 &
0.5$\pm$0.2 & 1.4$\pm$0.3 & 0.5$\pm$0.2 \\
\hline 130-140 & 4.3$\pm$1.2 & 3.4$\pm$0.9 & 1.7$\pm$0.4 &
1.3$\pm$0.3 & 0.4$\pm$0.2 & 0.6$\pm$0.2 \\
\hline 140-150 & 3.7$\pm$1.1 & 1.3$\pm$0.5 & 1.4$\pm$0.4 &
1.0$\pm$0.3 & 0.9$\pm$0.3 & 0.7$\pm$0.2\\
\hline 150-160 & 1.7$\pm$0.7 &   1.1$\pm$0.5 & 0.7$\pm$0.3 &
0.3$\pm$0.2 & 0.4$\pm$0.2 & 0.5$\pm$0.2\\
\hline 160-170 & 0.7$\pm$0.5  &  0.2$\pm$0.2 & 0.5$\pm$0.2 &
0.2$\pm$0.1 & 0.4$\pm$0.2 & 0.1$\pm$0.1 \\
\hline 170-180 & 1.0$\pm$0.6 & 0.2$\pm$0.2 & 0.4$\pm$0.2 &
0.5$\pm$0.2 & 0.1$\pm$0.1 & 0.2$\pm$0.1 \\
\hline
\end{tabular}
\end{center}

\vskip60pt

\begin{center}
{\bf T a b l e  2:  $^4$He({$\gamma$,p)$^3$H reaction}}
\begin{tabular}[t]{|c|c|c|c|c|c|c|}
\hline & \multicolumn{6}{|c|}{$E_{\gamma}$, MeV} \\ \cline{2-7}
$\theta_p^{\ast}$,deg & 22-23 & 23-24 &  24-25 & 25-26 & 26-27 &
27-28 \\
\hline 0-10 &   5.8$\pm$1.9 & 3.0$\pm$1.8 & 2.2$\pm$1.6 & 4.6$\pm$1.7 & 7.3$\pm$2.7 & 4.6$\pm$2.3 \\
\hline 10-20 &  17.3$\pm$3.3  &  20.3$\pm$4.5  &  22.3$\pm$5.0  &
17.2$\pm$3.4 & 17.6$\pm$4.3  &  23.1$\pm$5.2 \\
\hline 20-30 & 40.4$\pm$5.1  &  56.8$\pm$7.6  &  42.4$\pm$6.9  &
58.9$\pm$6.2 & 62.2$\pm$8.0  &  57.8$\pm$8.2 \\
\hline 30-40 & 75.7$\pm$7.0  &  96.4$\pm$9.9  &  81.4$\pm$9.5  &
76.7$\pm$7.1 & 105.7$\pm$10.5 & 78.6$\pm$9.5\\
\hline 40-50 & 103.9$\pm$8.2 &  135.9$\pm$11.7 & 104.8$\pm$10.8 &
109.8$\pm$8.5 & 141.9$\pm$12.1 & 120.3$\pm$11.8\\
\hline 50-60 & 148.1$\pm$9.7  & 156.2$\pm$12.6 & 131.5$\pm$12.1 &
134.9$\pm$9.4 & 161.6$\pm$12.9 & 164.2$\pm$13.8 \\
\hline 60-70 & 149.4$\pm$9.8 &  184.6$\pm$13.7 & 172.8$\pm$13.9 &
155.4$\pm$10.1 & 198.9$\pm$14.4 & 163.0$\pm$13.7 \\
\hline 70-80 & 162.9$\pm$10.2 & 195.8$\pm$14.1 & 170.6$\pm$13.8 &
182.5$\pm$11.0 & 226.9$\pm$15.3 & 197.7$\pm$15.1 \\
\hline 80-90 & 186.6$\pm$10.9 & 239.4$\pm$15.6 & 198.4$\pm$14.9 &
181.9$\pm$11.0 & 211.3$\pm$14.8 & 187.3$\pm$14.7 \\
\hline 90-100 & 175.1$\pm$10.6 & 229.3$\pm$15.2 & 188.4$\pm$14.5 &
164.7$\pm$10.4 & 195.8$\pm$14.2 & 188.5$\pm$14.8\\
\hline 100-110 & 155.8$\pm$10.0 & 193.7$\pm$14.0 & 169.5$\pm$13.7
& 163.4$\pm$10.4 & 185.4$\pm$13.9 & 133.0$\pm$12.4 \\
\hline 110-120 & 152.6$\pm$9.9 &  148.1$\pm$12.3 & 152.7$\pm$13.0
& 136.9$\pm$9.5 & 161.6$\pm$12.9 & 143.4$\pm$12.9 \\
\hline 120-130 & 119.3$\pm$8.7 &  144.0$\pm$12.1 & 133.8$\pm$12.2
& 113.8$\pm$8.7 & 126.4$\pm$11.4 & 91.3$\pm$10.3 \\
\hline 130-140 & 99.4$\pm$8.0  &  92.3$\pm$9.7  &  94.8$\pm$10.3 &
77.4$\pm$7.2 & 86.0$\pm$9.4  &  75.2$\pm$9.3 \\
\hline 140-150 & 62.2$\pm$6.3  &  58.8$\pm$7.7  &  58.0$\pm$8.0 &
58.9$\pm$6.2 & 70.4$\pm$8.5  &  55.5$\pm$8.0\\
\hline 150-160 & 38.5$\pm$5.0  &  29.4$\pm$5.5  &  37.9$\pm$6.5 &
21.2$\pm$3.7 & 31.1$\pm$5.7  &  27.8$\pm$5.7 \\
\hline 160-170 & 14.1$\pm$3.0  &  17.2$\pm$4.2  &  16.7$\pm$4.3 &
13.9$\pm$3.0 & 10.4$\pm$3.3  &  17.3$\pm$4.5 \\
\hline 170-180 & 4.5$\pm$1.7 & 7.1$\pm$2.7 & 5.6$\pm$2.5 &
3.3$\pm$1.5 & 5.2$\pm$2.3 & 6.9$\pm$2.8\\ \hline
\end{tabular}
\end{center}

\begin{center}
\begin{tabular}[t]{|c|c|c|c|c|c|c|}
\hline & \multicolumn{6}{|c|}{$E_{\gamma}$, MeV} \\ \cline{2-7}
$\theta_p^{\ast}$,deg & 28-29 & 29-30 &  30-31 & 31-32 & 32-33 &
33-34 \\ \hline 0-10 &   10.3$\pm$3.4 &   6.8$\pm$2.8 &
3.6$\pm$2.1 & 3.9$\pm$1.7 & 6.7$\pm$2.2 & 9.0$\pm$2.8 \\ \hline
10-20 & 24.1$\pm$5.3 &   27.1$\pm$5.5 &   25.0$\pm$5.4  &
22.4$\pm$4.2 & 23.0$\pm$4.1  &  17.1$\pm$3.9 \\ \hline 20-30 &
44.8$\pm$7.2  &  57.7$\pm$8.1  &  65.4$\pm$8.8  & 52.6$\pm$6.4 &
49.0$\pm$6.0  &  54.0$\pm$7.0 \\ \hline 30-40 & 74.7$\pm$9.3  &
106.3$\pm$11.0 & 89.1$\pm$10.3  & 75.1$\pm$7.6 & 86.8$\pm$8.0  &
63.8$\pm$7.6 \\ \hline 40-50 & 132.2$\pm$12.3 & 133.5$\pm$12.3 &
145.0$\pm$13.1 & 140.1$\pm$10.4 & 128.4$\pm$9.8 &  117.8$\pm$10.3\\
\hline 50-60 & 156.4$\pm$13.4 & 191.1$\pm$14.7 & 185.4$\pm$14.8 &
160.2$\pm$11.1 & 150.6$\pm$10.6 & 130.4$\pm$10.8 \\ \hline 60-70 &
178.2$\pm$14.3 & 193.4$\pm$14.8 & 173.5$\pm$14.4 & 160.2$\pm$11.1
& 172.1$\pm$11.3 & 147.5$\pm$11.5 \\ \hline 70-80 & 201.2$\pm$15.2
& 158.3$\pm$13.4 & 185.4$\pm$14.8 & 152.4$\pm$10.9 &
181.8$\pm$11.6 & 146.6$\pm$11.5 \\ \hline 80-90 &180.5$\pm$14.4 &
209.2$\pm$15.4 & 186.6$\pm$14.9 & 146.3$\pm$10.6 & 166.2$\pm$11.1
& 140.3$\pm$11.2 \\ \hline 90-100 & 163.3$\pm$13.7 &
166.3$\pm$13.7 & 158.0$\pm$13.7 & 147.0$\pm$10.7 & 151.4$\pm$10.6
& 115.1$\pm$10.2 \\ \hline 100-110 & 143.7$\pm$12.9 &
149.3$\pm$13.0 & 152.1$\pm$13.4 & 130.8$\pm$10.1 & 109.8$\pm$9.0 &
101.6$\pm$9.6 \\ \hline 110-120 & 131.1$\pm$12.3 & 132.3$\pm$12.2
& 108.1$\pm$11.3 & 107.6$\pm$9.1 & 102.4$\pm$8.7 &  93.5$\pm$9.2\\
\hline 120-130 &102.3$\pm$10.8 & 105.2$\pm$10.9 & 80.8$\pm$9.8  &
82.0$\pm$8.0 & 81.6$\pm$7.8  &  73.7$\pm$8.1 \\ \hline 130-140 &
64.4$\pm$8.6  &  76.9$\pm$9.3  &  76.1$\pm$9.5 & 67.3$\pm$7.2 &
59.4$\pm$6.6  &  45.9$\pm$6.4 \\ \hline 140-150 & 59.8$\pm$8.3  &
56.6$\pm$8.0  &  45.2$\pm$7.3 & 41.8$\pm$5.7 & 43.0$\pm$5.7  &
48.6$\pm$6.6 \\ \hline 150-160 & 42.5$\pm$7.0  &  26.0$\pm$5.4  &
26.1$\pm$5.6 & 34.8$\pm$5.2 & 18.6$\pm$3.7  &  28.8$\pm$5.1 \\
\hline 160-170 & 13.8$\pm$4.0  &  10.2$\pm$3.4  &  8.3$\pm$3.1 &
11.6$\pm$3.0  & 10.4$\pm$2.8  &  7.2$\pm$2.5\\ \hline 170-180 &
2.3$\pm$1.6 & 3.4$\pm$2.0 & 2.4$\pm$1.7 & 7.7$\pm$2.4 &
5.9$\pm$2.1 & 3.6$\pm$1.8 \\ \hline
\end{tabular}
\end{center}

\begin{center}
\begin{tabular}[t]{|c|c|c|c|c|c|c|}
\hline & \multicolumn{6}{|c|}{$E_{\gamma}$, MeV} \\ \cline{2-7}
$\theta_p^{\ast}$,deg & 34-35 & 35-36 &  36-37 & 37-38 & 38-39 &
39-40\\
\hline 0-10 &   11.1$\pm$3.3  &  4.5$\pm$2.0 & 7.1$\pm$2.7 &
6.5$\pm$2.7 & 3.7$\pm$2.1 & 8.6$\pm$3.3 \\
\hline 10-20 &  21.1$\pm$4.6 &   16.4$\pm$3.9  &  25.5$\pm$5.1  &
13.0$\pm$3.8 & 15.9$\pm$4.4  &  12.4$\pm$3.9 \\
\hline 20-30 & 39.2$\pm$6.3  &  58.2$\pm$7.3  &  42.8$\pm$6.6  &
34.7$\pm$6.1 & 50.1$\pm$7.8  &  48.2$\pm$7.7 \\
\hline 30-40 & 68.4$\pm$8.3  &  86.4$\pm$8.9  &  63.2$\pm$8.0  &
74.8$\pm$9.0 & 74.5$\pm$9.5  &  50.7$\pm$7.9\\
\hline 40-50 & 97.6$\pm$9.9  &  102.8$\pm$9.7  & 86.6$\pm$9.4  &
97.6$\pm$10.3 & 108.8$\pm$11.5 & 85.3$\pm$10.3\\
\hline 50-60 & 110.7$\pm$10.6 & 132.8$\pm$11.0 & 105.0$\pm$10.3 &
110.6$\pm$11.0 & 97.8$\pm$10.9 &  111.2$\pm$11.7 \\
\hline 60-70 & 123.8$\pm$11.2 & 148.3$\pm$11.6 & 122.3$\pm$11.2 &
104.1$\pm$10.6 & 103.9$\pm$11.3 & 87.7$\pm$10.4 \\
\hline 70-80 & 125.8$\pm$11.3 & 124.6$\pm$10.6 & 106.0$\pm$10.4 &
129.0$\pm$11.8 & 122,2$\pm$ 12.2 & 107.5$\pm$11.5 \\
\hline 80-90 & 133.8$\pm$11.6 & 133.7$\pm$11.0 & 130.4$\pm$11.5 &
111.7$\pm$11.0 & 92.9$\pm$10.7 &  86.5$\pm$10.3\\
\hline 90-100 & 106.7$\pm$10.4 & 165.6$\pm$12.3 & 94.8$\pm$9.8  &
91.1$\pm$9.9 & 88.0$\pm$10.4  & 81.6$\pm$10.0\\
\hline 100-110 & 85.5$\pm$9.3  &  109.2$\pm$10.0 & 81.5$\pm$9.1 &
70.5$\pm$8.7 & 74.5$\pm$9.5  &  76.6$\pm$9.7 \\
\hline 110-120 & 69.4$\pm$8.4  &  71.0$\pm$8.0  &  73.4$\pm$8.6 &
67.2$\pm$8.5 & 59.9$\pm$8.6  &  59.3$\pm$8.6\\
\hline 120-130 & 72.8$\pm$8.5  &  61.9$\pm$7.5  &  42.8$\pm$6.6 &
53.1$\pm$7.6 & 39.1$\pm$6.9  &  42.0$\pm$7.2 \\
\hline 130-140 & 52.3$\pm$7.3  &  51.9$\pm$6.9  &  37.7$\pm$6.2 &
55.3$\pm$7.7 & 25.7$\pm$5.6  &  30.9$\pm$6.2 \\
\hline 140-150 & 37.2$\pm$6.1  &  35.5$\pm$5.7  &  27.5$\pm$5.3 &
18.4$\pm$4.5 & 18.3$\pm$4.7  &  16.1$\pm$4.5 \\
\hline 150-160 & 36.2$\pm$6.0  &  22.7$\pm$4.5  &  9.2$\pm$3.1 &
11.9$\pm$3.6 & 25.7$\pm$5.6  &  12.4$\pm$3.9 \\
\hline 160-170 & 10.1$\pm$3.2  &  7.3$\pm$2.6 & 7.1$\pm$2.7 &
9.8$\pm$3.3 & 1.2$\pm$1.2 & 2.5$\pm$1.7\\
\hline 170-180 & 7.0$\pm$2.7 & 3.6$\pm$1.8 & 2.0$\pm$1.4 &
4.3$\pm$2.2 & 1.2$\pm$1.2 & 3.7$\pm$2.1\\
\hline
\end{tabular}
\end{center}

\vskip20pt

\begin{center}
\begin{tabular}[t]{|c|c|c|c|c|c|c|}
\hline & \multicolumn{6}{|c|}{$E_{\gamma}$, MeV} \\ \cline{2-7}
$\theta_p^{\ast}$,deg &40-41 & 41-42 &  42-43 & 43-44 & 44-46 &
46-48\\
\hline 0-10 &   3.4$\pm$2.4 & 4.5$\pm$2.6 & 3.5$\pm$2.5 &
4.6$\pm$2.7 & 6.6$\pm$3.8 & 4.1$\pm$1.8 \\
\hline 10-20 &  8.4$\pm$3.8 & 16.4$\pm$5.0  &  10.6$\pm$4.3  &
20.1$\pm$5.6 & 13.3$\pm$5.4  &  9.0$\pm$2.7 \\
\hline 20-30 & 47.1$\pm$8.9  &  28.4$\pm$6.5 &   24.7$\pm$6.6  &
40.3$\pm$7.9 & 24.3$\pm$7.3  &  27.7$\pm$4.8 \\
\hline 30-40 & 52.2$\pm$9.4  &  52.3$\pm$8.8  &  47.7$\pm$9.2  &
49.6$\pm$8.8 & 44.2$\pm$9.9  &  43.2$\pm$5.9\\
\hline 40-50 & 52.2$\pm$9.4  &  85.2$\pm$11.3 &  61.8$\pm$10.4 &
66.6$\pm$10.2 & 64.1$\pm$11.9 &  55.4$\pm$6.7 \\
\hline 50-60 & 77.4$\pm$11.4  & 85.2$\pm$11.3 &  88.3$\pm$12.5  &
71.3$\pm$10.5 & 50.9$\pm$10.6 &  73.3$\pm$7.7 \\
\hline 60-70 & 87.5$\pm$12.1  & 80.7$\pm$11.0  & 86.5$\pm$12.4  &
93.0$\pm$12.0 & 73.0$\pm$12.7  & 79.0$\pm$8.0 \\
\hline 70-80 & 67.3$\pm$10.6 &  88.2$\pm$11.5  & 77.7$\pm$11.7  &
97.6$\pm$12.3 & 70.8$\pm$12.5  & 76.6$\pm$7.9 \\
\hline 80-90 & 94.3$\pm$12.6  & 79.2$\pm$10.9 &  70.6$\pm$11.2  &
71.3$\pm$10.5 & 64.1$\pm$11.9 &  68.4$\pm$7.5\\
\hline 90-100 & 72.4$\pm$11.0 &  62.8$\pm$9.7  &  56.5$\pm$10.0 &
85.2$\pm$11.5 & 50.9$\pm$10.6  & 67.6$\pm$7.4\\
\hline 100-110 & 65.6$\pm$10.5 &  67.3$\pm$10.0 &  51.2$\pm$9.5 &
55.8$\pm$9.3 & 35.4$\pm$8.8  &  57.0$\pm$6.8 \\
\hline 110-120 & 50.5$\pm$9.2  &  37.4$\pm$7.5  &  35.3$\pm$7.9 &
48.0$\pm$8.6 & 46.4$\pm$10.1  & 34.2$\pm$5.3 \\
\hline 120-130 & 45.4$\pm$8.7  &  41.9$\pm$7.9  &  45.9$\pm$9.0 &
49.6$\pm$8.8 & 31.0$\pm$8.3  &  27.7$\pm$4.8 \\
\hline 130-140 & 26.9$\pm$6.7  &  43.4$\pm$8.1  &  26.5$\pm$6.8 &
20.1$\pm$5.6 & 13.3$\pm$5.4  &  23.6$\pm$4.4 \\
\hline 140-150 & 15.1$\pm$5.0  &  22.4$\pm$5.8   & 17.7$\pm$5.6 &
27.9$\pm$6.6 & 15.5$\pm$5.9  &  16.3$\pm$3.6\\
\hline 150-160 & 16.8$\pm$5.3  &  10.5$\pm$4.0  &  10.6$\pm$4.3 &
7.7$\pm$3.5 & 2.2$\pm$2.2 & 10.6$\pm$2.9 \\
\hline 160-170 & 3.4$\pm$2.4 & 4.5$\pm$2.6 & 8.8$\pm$3.9 &
4.6$\pm$2.7 & 2.2$\pm$2.2 & 6.5$\pm$2.3\\
\hline 170-180 & 3.4$\pm$2.4 & 3.0$\pm$2.1 & 5.3$\pm$3.1 &
6.2$\pm$3.1 & 4.4$\pm$3.1 & 4.9$\pm$2.0\\
\hline
\end{tabular}
\end{center}

\begin{center}
\begin{tabular}[t]{|c|c|c|c|c|c|c|}
\hline & \multicolumn{6}{|c|}{$E_{\gamma}$, MeV} \\ \cline{2-7}
$\theta_p^{\ast}$,deg & 58-60 & 60-62 &  62-64 & 64-66 & 66-68 &
68-70\\
\hline 0-10 &   0.0$\pm$0.9 & 0.6$\pm$0.6 & 0.6$\pm$0.6 & 0.7$\pm$0.7 & 0.0$\pm$0.6 & 1.1$\pm$0.8 \\
\hline 10-20 &  6.8$\pm$2.4 & 5.8$\pm$1.8 & 8.9$\pm$2.4 & 5.8$\pm$2.0 & 1.2$\pm$0.8 & 3.8$\pm$1.5\\
\hline 20-30 & 10.3$\pm$3.0  &  16.9$\pm$3.1 &   9.5$\pm$2.5 & 13.8$\pm$3.2  &  10.0$\pm$2.4 & 11.0$\pm$2.5 \\
\hline 30-40 & 18.0$\pm$3.9  &  23.3$\pm$3.7  &  29.9$\pm$4.4  &  18.1$\pm$3.6  &  18.8$\pm$3.3  &  8.8$\pm$2.2\\
\hline 40-50 & 35.1$\pm$5.5  &  26.2$\pm$3.9  &  29.9$\pm$4.4  &  25.3$\pm$4.3  &  20.6$\pm$3.5  &  14.8$\pm$2.9 \\
\hline 50-60 & 35.1$\pm$5.5  &  27.9$\pm$4.0   & 28.6$\pm$4.3  &  24.6$\pm$4.2  &  22.9$\pm$3.7  &  21.4$\pm$3.4 \\
\hline 60-70 & 37.6$\pm$5.7  &  33.1$\pm$4.4  &  22.3$\pm$3.8  &  33.3$\pm$4.9  &  33.5$\pm$4.4  &  19.2$\pm$3.3 \\
\hline 70-80 & 36.8$\pm$5.6  &  29.7$\pm$4.2  &  29.9$\pm$4.4   & 33.3$\pm$4.9  &  32.3$\pm$4.4  &  34.6$\pm$4.4 \\
\hline 80-90 & 30.8$\pm$5.1  &  25.0$\pm$3.8   & 20.4$\pm$3.6  &  34.0$\pm$5.0  &  25.8$\pm$3.9   & 22.0$\pm$3.5\\
\hline 90-100 & 32.5$\pm$5.3  &  23.3$\pm$3.7  &  24.8$\pm$4.0  &  22.4$\pm$4.0   & 9.4$\pm$2.3 & 15.9$\pm$3.0\\
\hline 100-110 & 17.1$\pm$3.8  &  14.5$\pm$2.9  &  11.5$\pm$2.7  &  17.4$\pm$3.5   & 9.4$\pm$2.3 & 17.0$\pm$3.1\\
\hline 110-120 & 18.0$\pm$3.9  &  16.9$\pm$3.1  &  15.9$\pm$3.2  &  10.1$\pm$2.7  &  11.2$\pm$2.6  &  11.0$\pm$2.5 \\
\hline 120-130 & 12.0$\pm$3.2  &  10.5$\pm$2.5  &  8.3$\pm$2.3 & 7.2$\pm$2.3 & 7.6$\pm$2.1 & 6.0$\pm$1.8 \\
\hline 130-140 & 12.8$\pm$3.3  &  12.2$\pm$2.7  &  5.1$\pm$1.8 & 5.8$\pm$2.0 & 5.3$\pm$1.8 & 7.7$\pm$2.1\\
\hline 140-150 & 6.8$\pm$2.4 & 4.7$\pm$1.6 & 4.5$\pm$1.7 & 9.4$\pm$2.6 & 2.3$\pm$1.2 & 4.4$\pm$1.6\\
\hline 150-160 & 6.0$\pm$2.3 & 4.7$\pm$1.6 & 1.9$\pm$1.1 & 4.3$\pm$1.8 & 2.9$\pm$1.3 & 1.1$\pm$0.8\\
\hline 160-170 & 0.9$\pm$0.9 & 1.2$\pm$0.8 & 1.9$\pm$1.1 & 1.4$\pm$1.0 & 1.8$\pm$1.0 & 2.7$\pm$1.2\\
\hline 170-180 & 0.0$\pm$0.9 & 0.6$\pm$0.6 & 0.6$\pm$0.6 & 0.7$\pm$0.7 & 2.3$\pm$1.2 & 0.5$\pm$0.5\\
\hline
\end{tabular}
\end{center}

\begin{center}
\begin{tabular}[t]{|c|c|c|c|c|c|c|}
\hline & \multicolumn{6}{|c|}{$E_{\gamma}$, MeV} \\ \cline{2-7}
$\theta_p^{\ast}$,deg &70-72 & 72-74 &  74-76 & 76-78 & 78-80 &
80-85\\
\hline 0-10 &   0.5$\pm$0.5 & 1.1$\pm$0.8 & 1.0$\pm$0.7 & 0.0$\pm$0.4 & 0.9$\pm$0.7 & 0.3$\pm$0.3 \\
\hline 10-20 &  3.8$\pm$1.4 & 4.5$\pm$1.6 & 3.0$\pm$1.2 & 3.7$\pm$1.2 & 4.3$\pm$1.4 & 4.6$\pm$1.2 \\
\hline 20-30 & 8.7$\pm$2.2 & 10.0$\pm$2.4  &  8.5$\pm$2.1 & 8.2$\pm$1.8 & 9.0$\pm$2.1 & 5.9$\pm$1.4 \\
\hline 30-40 & 16.9$\pm$3.0  &  11.2$\pm$2.5  &  13.5$\pm$2.6  &  9.1$\pm$1.9 & 11.8$\pm$2.4  &  7.9$\pm$1.6\\
\hline 40-50 & 16.9$\pm$3.0  &  16.2$\pm$3.0  & 15.0$\pm$ 2.7  &  14.0$\pm$2.4  &  19.4$\pm$3.0  &  12.4$\pm$2.0 \\
\hline 50-60 & 20.7$\pm$3.4  &  21.2$\pm$3.4  &  20.5$\pm$3.2  &  17.3$\pm$2.7  &  17.0$\pm$2.8  &  11.5$\pm$1.9 \\
\hline 60-70 & 21.8$\pm$3.4  &  24.0$\pm$3.7  &  21.0$\pm$3.2  &  19.0$\pm$2.8  &  24.1$\pm$3.4  &  17.0$\pm$2.4 \\
\hline 70-80 & 24.0$\pm$3.6  &  22.3$\pm$3.5  &  17.5$\pm$3.0   & 13.2$\pm$2.3   & 18.5$\pm$3.0  &  11.8$\pm$2.0 \\
\hline 80-90 & 18.5$\pm$3.2  &  21.8$\pm$3.5  &  22.0$\pm$3.3  &  13.6$\pm$2.4  &  17.0$\pm$2.8  &  9.2$\pm$1.7\\
\hline 90-100 & 16.9$\pm$3.0  &  19.0$\pm$3.3  &  15.0$\pm$2.7  &  12.8$\pm$2.3  &  17.5$\pm$2.9  &  9.5$\pm$1.8\\
\hline 100-110 & 14.7$\pm$2.8  &  14.0$\pm$2.8  &  9.0$\pm$2.1 & 7.4$\pm$1.7 & 9.5$\pm$2.1 & 6.9$\pm$1.5 \\
\hline 110-120 & 14.2$\pm$2.8  &  6.1$\pm$1.9 & 12.0$\pm$2.5 &  8.7$\pm$1.9 & 11.8$\pm$2.4  &  6.9$\pm$1.5\\
\hline 120-130 & 9.3$\pm$2.2 & 12.8$\pm$2.7  &  8.0$\pm$2.0 & 7.4$\pm$1.7 & 3.3$\pm$1.3 & 6.9$\pm$1.5 \\
\hline 130-140 & 5.4$\pm$1.7 & 9.5$\pm$2.3 & 7.0$\pm$1.9 & 6.2$\pm$1.6 & 4.3$\pm$1.4 & 4.6$\pm$1.2 \\
\hline 140-150 & 3.3$\pm$1.3 & 6.7$\pm$1.9 & 5.5$\pm$1.7 & 4.9$\pm$1.4 & 1.9$\pm$0.9 & 3.3$\pm$1.0\\
\hline 150-160 & 2.7$\pm$1.2  &  2.8$\pm$1.2 & 5.5$\pm$1.7 & 3.3$\pm$1.2 & 1.4$\pm$0.8 & 1.3$\pm$0.7 \\
\hline 160-170 & 3.3$\pm$1.3 & 1.7$\pm$1.0 & 0.5$\pm$0.5 & 2.9$\pm$1.1 & 2.8$\pm$1.2 & 1.3$\pm$0.7\\
\hline 170-180 & 0.0$\pm$0.5 & 1.7$\pm$1.0 & 0.5$\pm$0.5 & 0.8$\pm$0.6 & 0.5$\pm$0.5 & 0.0$\pm$0.3\\
\hline
\end{tabular}
\end{center}

\begin{center}
\begin{tabular}[t]{|c|c|c|c|c|c|c|c|c|}
\hline & \multicolumn{7}{|c|}{$E_{\gamma}$, MeV} \\ \cline{2-8}
$\theta_p^{\ast}$,deg &85-90 & 90-95 &  95-100 & 100-110 & 110-120
& 120-130 & 130-140\\
\hline 0-10 &   0.6$\pm$0.5 & 0.9$\pm$0.5 & 0.0$\pm$0.2 & 0.8$\pm$0.3 & 0.2$\pm$0.1 & 0.1$\pm$0.1 & 0.1$\pm$0.1 \\
\hline 10-20 &  2.6$\pm$0.9 & 1.6$\pm$0.7 & 2.7$\pm$0.8 & 1.1$\pm$0.4 & 0.9$\pm$0.3 & 0.3$\pm$0.1 & 0.4$\pm$0.1 \\
\hline 20-30 & 6.8$\pm$1.5 & 5.0$\pm$1.2 & 5.4$\pm$1.1 & 2.5$\pm$0.6 & 1.6$\pm$0.4 & 1.1$\pm$0.3 & 1.4$\pm$0.3 \\
\hline 30-40 & 6.5$\pm$1.4 & 7.5$\pm$1.5 & 6.8$\pm$1.2 & 4.1$\pm$0.7 & 2.6$\pm$0.5 & 2.1$\pm$0.4 & 1.8$\pm$0.3\\
\hline 40-50 & 10.0$\pm$1.8  &  9.9$\pm$1.8 & 7.4$\pm$1.3 & 5.3$\pm$0.8 & 3.5$\pm$0.6 & 2.1$\pm$0.4 & 1.5$\pm$0.5 \\
\hline 50-60 & 16.5$\pm$2.3  &  15.5$\pm$2.2  &  7.9$\pm$1.3 & 5.6$\pm$0.9 & 3.4$\pm$0.6 & 2.9$\pm$0.4 & 2.6$\pm$0.4 \\
\hline 60-70 & 12.9$\pm$2.0  &  13.4$\pm$2.0  &  9.7$\pm$1.5 & 5.9$\pm$0.9 & 3.7$\pm$0.6 & 2.4$\pm$0.4 & 2.2$\pm$0.3 \\
\hline 70-80 & 13.2$\pm$2.1  &  13.0$\pm$2.0   & 11.3$\pm$1.6  &  4.5$\pm$0.8 & 3.9$\pm$0.6 & 3.1$\pm$0.4 & 2.0$\pm$0.3 \\
\hline 80-90 & 12.0$\pm$2.0  &   10.9$\pm$1.8   & 7.0$\pm$1.3 & 4.1$\pm$0.7 & 3.6$\pm$0.6 & 1.9$\pm$0.3 & 1.7$\pm$0.3\\
\hline 90-100 & 9.0$\pm$1.7 & 7.1$\pm$1.5 & 6.1$\pm$1.2 & 3.6$\pm$0.7 & 2.9$\pm$0.5 & 1.7$\pm$0.3 & 1.4$\pm$0.3\\
\hline 100-110 & 8.4$\pm$1.6 & 6.8$\pm$1.5 & 5.0$\pm$1.1 & 2.4$\pm$0.6 & 1.6$\pm$0.4 & 1.3$\pm$0.3 & 1.0$\pm$0.2 \\
\hline 110-120 & 6.5$\pm$1.4 & 7.8$\pm$1.6 & 3.2$\pm$0.8 & 3.5$\pm$0.7 & 1.7$\pm$0.4 & 0.9$\pm$0.2 & 0.8$\pm$0.2 \\
\hline 120-130 & 3.6$\pm$1.1 & 6.5$\pm$1.4 & 5.2$\pm$1.1 & 1.7$\pm$0.5 & 0.8$\pm$0.3 & 0.6$\pm$0.2 & 0.5$\pm$0.2 \\
\hline 130-140 & 5.2$\pm$1.3 & 2.5$\pm$0.9 & 1.8$\pm$0.6 & 1.3$\pm$0.4 & 1.2$\pm$0.3 & 0.8$\pm$0.2 & 0.4$\pm$0.1 \\
\hline 140-150 & 3.9$\pm$1.1 & 3.4$\pm$1.0 & 2.0$\pm$0.7 & 1.9$\pm$0.5 & 1.0$\pm$0.3 & 0.5$\pm$0.2 & 0.4$\pm$0.1\\
\hline 150-160 & 1.0$\pm$0.6 & 1.2$\pm$0.6 & 0.9$\pm$0.5 & 0.7$\pm$0.3 & 0.4$\pm$0.2 & 0.2$\pm$0.1 & 0.5$\pm$0.2\\
\hline 160-170 & 0.6$\pm$0.5 & 0.6$\pm$0.4 & 0.5$\pm$0.3 & 0.1$\pm$0.1 & 0.3$\pm$0.2 & 0.1$\pm$0.1 & 0.3$\pm$0.1\\
\hline 170-180 & 0.3$\pm$0.3 & 1.2$\pm$0.6 & 0.7$\pm$0.4 & 0.1$\pm$0.1 & 0.2$\pm$0.1 & 0.4$\pm$0.1 & 0.1$\pm$0.1\\
\hline
\end{tabular}
\end{center}

\end{document}